\documentstyle[11pt]{article}
\begin{document} 
\makeatletter
\@addtoreset{equation}{section}
\makeatother
\renewcommand{\theequation}{\thesection.\arabic{equation}}
\baselineskip 15pt

\title{\bf A Quantum Approach To Static Games Of Complete Information
\footnote{Work supported in part by Istituto Nazionale di Fisica Nucleare,
 Sezione di Trieste, Italy}}
\author{Luca Marinatto\footnote{e-mail: marinatto@ts.infn.it}\\
{\small Department of Theoretical Physics of the University of Trieste, and}\\
{\small Istituto Nazionale di Fisica Nucleare, Sezione di Trieste, Italy.}\\
and \\
\\ Tullio Weber\footnote{e-mail: weber@ts.infn.it }\\
{\small Department of Theoretical Physics of the University of Trieste, and}\\
{\small Istituto Nazionale di Fisica Nucleare, Sezione di Trieste, Italy.}}

\date{}

\maketitle

\begin{abstract}
We extend the concept of a classical two-person static game to the quantum 
domain, by giving an Hilbert structure to the space of classical strategies
 and studying the {\em Battle of
 the Sexes} game. We show that the introduction of entangled strategies leads
 to a unique solution of this game.\\

  

\end{abstract}


\section{Introduction.}

Classical Game Theory concerns the study of multiperson decision problems,
 where two or more individuals make rational decisions that will influence one
 another's welfare.
The kind of problems the theory faces off arise frequently in economics and
 in social sciences, ranging from the conflicts of firms in the market to the
foreign policy of nations.
As far as physics is concerned, there is an intimate connection between
the theory of games and quantum communication theory, where the task of two
 distant players is to obtain as much information as possible in a given
physical situation.

Modern Game Theory was first developed in 1944 in the seminal book of Von
 Neumann and Morgenstern \cite{ref1}, but a better comprehension of its formal
 structure and the major mathematical results were achieved in the following 
 years at Princeton, mainly due to the contribution of the young brilliant 
mathematician John Nash.
The theory tries to understand the birth and the development of conflictual
 or cooperational behaviours between rational and intelligent decision makers,
 by analyzing simple but paradigmatic problems, which retain 
 the crucial features exhibited by the real situations in every day life.
It is worthwhile to stress that in the theoretical language a
 game refers to any situation where two or more persons, called
 players, are involved in a cooperation or a competition among themselves to
 reach a final state recognized as the best goal they can obtain from a
 cooperative or an individual point of view.
 It is clear that this general definition encompasses the usual meaning given
 to a game, i.e. that of a play ($\,$like chess and
 draughts$\,$), but it applies to bargaining or trading processes too.
The persons who play the game are assumed to think, and consequently act, in a
rational way, by trying to maximize the expected values of some functions,
 the payoff functions, which depend on the choices of all the players and
represent their gain at the end of the game.     
They are also considered as intelligent beings, in the sense that each
 player knows about the game everything that we know: therefore they can
 make the same inferences about the development of the game that we are
able to do.
          
The purpose of this paper is to try to formalize the concept of  Quantum Game
($\,$as opposed to  Classical Game$\,$), by giving a formal quantum structure 
to the usual mathematical tools commonly used in analysing decision problems.
We have taken inspiration from the paper by J. Eisert, M. Wilkens
 and M. Lewenstein~\cite{ref2}, but the formal structure of the Quantum
Game Theory we are trying to develop differs completely from their approach, 
though holding true that we agree about the special role played
by entanglement.

The passage from a classical to a quantum domain will be achieved by using
an Hilbert space of strategies, instead of a discrete set of them, then 
allowing the possibility of the existence of linear superpositions between
classical strategies: this will naturally enlarge the possible strategic
 choices offered to each player from a numerable to a continuous set.

As it will be evident from the analysis of one of the famous and paradigmatic
 classical games ($\,$the Battle of the Sexes$\,$), the use of
a quantum formalism will exhibit a novel feature, the appearance in 
{\it entangled strategies} of a Nash Equilibrium representing
 the unique solution of the game -- in contrast with what happens in the
 classical case, where the theory is not able to make any unique prediction.

In sections 2 and 3 the basic concepts of Classical Game Theory are enunciated
 and applied to the study of the forementioned game.
In section 4 we try to define what is meant by a
 quantum strategy, as opposed to a classical one, and in sections 5 and 6 we 
show how it could be exploited by each player to find out the Nash equilibria
 in the case of factorizable Quantum Strategies. In section 7 we extend the
 method to the entangled strategies, showing that a unique solution arises
 for the Battle of the Sexes game.


\section{Classical Game Theory}

In this chapter we want to provide an introductive view of the basic
 principles of the theory of the Static Games of Complete Information.
 Simple decision problems, in which both players simultaneously choose their
 actions ($\,$Static Game$\,$) and receive a payoff depending on their mutual
 choices, are contained within this restricted class of games.
 Moreover, each player knows perfectly the values of the payoff functions of
 all his opponents ($\,$Complete Information$\,$).
Games which do not belong to this class are Dynamic Games ($\,$the
 players play sequential moves by turn, like in chess$\,$) or Games of
Incomplete Information ($\,$some player does not know exactly the payoff
 functions of his opponents, like in a auction$\,$).   

\newpage  

Every {\bf Normal Form Representation} of our games must specify: 
 \begin{enumerate}
 \item the number of players $i=1\dots N$; 
 \item the discrete set of strategies $\{ s \}$, which
  constitutes the {\it strategic space} ${\cal{S}}_{i}$ available to each
 player;
 \item the payoff functions ${\$}_{i}={\$}_{i}(s_{1}, s_{2}\dots s_{N})$,
 which assign the i-th player a real number depending on the strategies
 chosen by his opponents.
\end{enumerate}

These functions quantify the benefit that the i-th player gains
if the strategies $(s_{1}, s_{2}\dots s_{N})$ are effectively played, and the
 purpose of each decision maker consists in trying to maximize, in a sense
 depending on the context of the game, this real number. 

The theory of games deals therefore with the forecast of which will be the
 most rational development of a given game: as we will see later, it will
 not necessarily be the most appealing one for each player, but the one that 
 each player is forced to play in order to avoid major losses.

In order to simplify the notation, we will denote from now on by $s_{i}$ the 
generic strategy belonging to the strategic space ${\cal{S}}_{i}$ of the
 i-th player.   
Each player chooses his own strategy simultaneously to the others,
($\,$i.e. without knowing what the other players  are going to do, but knowing
 what they might do$\,$) and has a complete information about the consequences
 of all the possible choices.
  
Having settled the common framework of all the Static Games, it is useful to
 introduce the notion of a strictly dominated strategy. 
 Strategy $s_{i}$ is {\bf strictly dominated} by strategy 
 $\tilde{s_{i}}$, if:

 \begin{equation}
 {\$}_{i}(s_{1},\ldots, s_{i}, \ldots, s_{N}) <  {\$}_{i}(s_{1},\ldots,
 \tilde{s_{i}},  \ldots, s_{N}), 
 \end{equation}

  for every set of strategies $(s_{1},\ldots, s_{i-1}, s_{i+1},\ldots, s_{N})$
 of the other players. 

This concept suggests a first kind of solution for a given game, where the
 word solution refers to the ensemble of the $N$ strategies which are actually
 played.
In fact, making the reasonable assumption that rational players do not play
 strictly dominated strategies ($\,$as we have said before, they pursue the
 goal of maximizing their benefits, whatever are the strategies of their
 opponents$\,$), we could find out the best strategy for each player
 by simply eliminating all the dominated ones.
The strategy obtained by such a procedure, which is termed {\it Iterated
 Elimination of Strictly Dominated Strategies}, represents the best action
 each player could reasonable perform against his opponent's decisions.
Therefore the set of strategies chosen in such a way constitutes a possible
 ($\,$rational$\,$) solution of the game.
 
Unfortunately, this appealing procedure produces no prediction at all
 for certain classes of problems, where no strategy survives the process of
 elimination. In such a case it is not clear 
 which action could be considered a rational and optimal decision.
   
 In order to illustrate the concepts we have just introduced, we analyse a
 simple two-person static game of complete information,
 which is known as the Battle of the Sexes.
 In the usual exposition of the game a woman, Alice, and a man,
 Bob, are trying to decide where to spend the saturday evening:
 Alice would be happy to attend the Opera, while Bob would like to watch the
 football match at the Television, and both of them would be happier to stay
 together rather than far apart. 
 We can represent this simple game in a normal form by denoting by
 O (Opera) and T (Television) the two strategies which constitute
 the common strategic space $\cal{S}$
 and we can represent their payoff functions by means of a bimatrix like the
 following one:

\begin{picture}(1000,140)
 \put(150,0){\framebox(100,80)}
 \put(150,40){\line(100,0){100}}
 \put(200,0){\line(0,80){80}}
 \put(165,15){($\gamma,\gamma$)}
 \put(215,15){($\beta,\alpha$)}
 \put(165,55){($\alpha,\beta$)}
 \put(215,55){($\gamma,\gamma$)}
 \put(130,15){\large\bf {T}}
 \put(130,55){\large\bf {O}}
 \put(170,90){\large\bf {O}}
 \put(220,90){\large\bf {T}}
 \put(70,35){\large\bf {Alice}}
 \put(185,110){\large\bf {Bob}}
\end{picture}
\vspace{0.5cm}

where $\alpha, \beta, \gamma$ are the values assumed by the payoff functions
 of Alice and Bob in correspondence to the possible choice of strategies: if,
 for example, both players decide to go to the Opera ($\,$i.e., if the couple
 of strategies $(O,O)$ is played$\,$), Alice gains ${\$}_{A}(O,O)=\alpha$ and
 Bob gains ${\$}_{B}(O,O)=\beta$.
In order to satisfy the preferences of the two players mentioned
 before, the condition $\alpha > \beta > \gamma$ must be imposed.
This constraint condition guarantees that we are dealing with a proper 
form of the Battle of the Sexes game, with the right options for Alice and Bob.

This is a peculiar game, where strictly dominated strategies do not exist
($\,$in fact according to Alice, for example, strategy O is better than T
if Bob decides to play O, and it is worse if Bob decides to play T$\,$):
 neither of the players can rationally eliminate one of his strategies and
 prefer the other.

To try to overcome this unpleasant situation, which is not peculiar of the
 Battle of the Sexes game, one is induced to resort in general to the
 concept of Nash Equilibrium, which permits it to get possible
 and rational solutions for the kind of games we are referring to ($\,$it
 is worthwhile remembering that the word {\em solution} corresponds to a
 set of strategies the rational players will surely play$\,$).
In a N-player normal form representation game, the set of strategies
$(s_{1}^{\star},s_{2}^{\star},\ldots,s_{N}^{\star})$ constitute a {\bf Nash
Equilibrium} if, for each player $i$, the strategy $s_{i}^{\star}$ satisfies
the following equation:

\begin{equation}
\label{Nashequlibrium}
 {\$}_{i}(s_{1}^{\star},\ldots,s_{i-1}^{\star},s_{i}^{\star},s_{i+1}^{\star},
\ldots,s_{N}^{\star}) \geq  
 {\$}_{i}(s_{1}^{\star},\ldots,s_{i-1}^{\star},s_{i},s_{i+1}^{\star},
\ldots,s_{N}^{\star}),
\end{equation}

for every strategy $s_{i}$ belonging to the i-th strategic space
 ${\cal{S}}_{i}$. 

Therefore a Nash Equilibrium corresponds to a set of strategies, each one 
 representing the best choice for each single player if all his opponents
 take their best decision too.
It exhibits the appealing property of being {\em strategically stable}, because
 no single player has the incentive to deviate unilaterally from his strategy
without suffering some loss in his own payoff.
The motivation leading to this new concept is based on the following
 properties:

\begin{enumerate}
\item it generalizes the previous and inadequate concept of {\em solution of
 a game}, for it can be easily proved that every set of 
 strategies surviving the process of elimination of strictly dominated
strategies constitutes the unique Nash Equilibrium of the game -- the converse
statement being not necessarily true.
 We will in particular see that the Battle of the Sexes possesses two Nash
 Equilibria in pure strategies ($\,$ representing two possible predictions
for the development of the game $\,$), while the process of elimination of
strategies cannot give any result;
\item it is, as already said, a {\em stable} equilibrium, in the sense that
no player could gain better by unilaterally deviating from his predicted
strategy. 
\end{enumerate}

Returning back to the analysis of our game and in order to find out which is
 the pair of strategies which satisfy the Nash-Equilibrium conditions
 (\ref{Nashequlibrium}), we must solve the two following coupled inequalities
 in the unknown strategies $s_{A}^{\star}$ and $s_{B}^{\star}$, where the
 letters $A$ and $B$ stand obviously for Alice and Bob:

\begin{equation}
\left\{
  \begin{array}{ll}
  {\$}_{A}(s_{A}^{\star},s_{B}^{\star}) \geq {\$}_{A}(s_{A},s_{B}^{\star})
  \:\:\:\:\:\:\:\:\:\: \forall \:\:s_{A}\in {\cal{S}},  \\
  \\
  {\$}_{B}(s_{A}^{\star},s_{B}^{\star}) \geq {\$}_{B}(s_{A}^{\star},s_{B})
  \:\:\:\:\:\:\:\:\:\: \forall \:\:s_{B}\in {\cal{S}}. 
\end{array}
\right.
\end{equation}

It is easy to verify that two couples of strategies satisfy these conditions
for the Battle of the Sexes Game: there exist therefore two Nash Equilibria,
 corresponding to the choices $(s^{\star}_{A}=O\,,\,s^{\star}_{B}=O\,)$ or 
$(s^{\star}_{A}=T\,,\,s^{\star}_{B}=T\,)$.
These strategies are strategically stable exactly in the sense given above.

The appearance of two possible solutions of the game in this peculiar
 case is rather discomforting ($\,$being the first one more favourable to
 Alice, while the second is favourable to Bob$\,$), even if this is better
 than having no solution, as it was the case before the Nash Equilibrium was
 introduced.
However, as we will show in the next sections, this problem disappears
in a quantum context, where we will be able to exhibit ($\,$in peculiar
 situations to be specified$\,$) only one solution of the game.


\section{Mixed Classical Strategies}

Although the notion of a Nash Equilibrium represents a very clever new concept
 of solution for a game, it is still possible to exhibit simple games not
 possessing Nash Equilibria, this fact leading to a situation that does not
 display any rational prediction about the development of the game.
This usually happens in games where each player would like to outguess
the decisions taken by his opponents ($\,$like in poker or in a battle$\,$) in
 order to gain an advantage over them.
 To solve this new kind of games we must resort to the introduction
 of the concept of the {\bf Mixed Classical Strategies}, as opposed to the
 Pure ones presented before: in a normal form representation game, a Mixed
 Strategy for the i-th player is a probability distribution which assigns a
 probability $p_{i}^{\alpha}$ to each pure strategy $s_{i}^{\alpha}$, where
 $\alpha$ runs over all the possible strategies belonging to the strategic
 space ${\cal{S}}_{i}$ of the i-th player. 

This means that each player chooses one of the possible strategies by resorting
 to chance, according to the probability distribution which could assure him
 the best result.

This definition generalizes the notion of strategy given before,
 since a pure strategy can always be seen as a mixed one with probability
 distribution which is equal to one for a particular strategy and
 zero for all the others. According to Harsanyi ~\cite{ref3}, it is possible
 to interpret such a mixed strategy in terms of the uncertainty of each player
 about what another player will do.

The introduction of a probability distribution in the space of strategies
 forces one to modify the concept of the payoff function for the i-th player,
 leading to an {\bf expected payoff function} $\bar{\$}_{i}$:

\begin{equation}
\label{expected}
\bar{\$}_{i}(\left\{ p_{1} \right\},\left\{ p_{2} \right\},\dots,\left\{ p_{N}
 \right\}) =
\sum_{\alpha,\beta,\dots,\gamma} p_{1}^{\alpha}p_{2}^{\beta}\ldots
 p_{N}^{\gamma}\, {\$}_{i}(s_{1}^{\alpha},s_{2}^{\beta},\ldots,s_{N}^{\gamma})
\end{equation}

where $\left\{ p_{i} \right\}$ is the set of the probabilities associated to
 the strategies of the space ${\cal S}_{i}$.
 
We note that $\bar{\$}_{i}$ is no longer a function of the strategies,
 depending now on their probability distributions and represents an average
 gain for the i-th player.

The usefulness in dealing with mixed strategies resides in a remarkable result
 proved by Nash~\cite{ref4}, stating that in every N-player normal form
 game with finite strategic spaces there always exists at least one Nash
 Equilibrium, in pure or mixed strategies.   
Therefore, one is led to modify slightly the definition of Nash Equilibrium in
 order to incorporate mixed strategies too.
In the two-player normal form games with two strategies, like the Battle of
 the Sexes, we are concerned with
 the probabilities $\left\{ p_{1}^{1}, p_{1}^{2}= 1- p_{1}^{1} \right\}$ for
 the first player, and $\left\{ p_{2}^{1}, p_{2}^{2}= 1- p_{2}^{1} \right\}$
for the second one.

Since only two of them are independent, we put $p_{1}^{1}=p$ and $p_{2}^{1}=q$,
writing $\bar{\$_{i}}(\left\{ p_{1} \right\},\left\{ p_{2} \right\})\equiv
\bar{\$_{i}}(p,q)$. Using this simplified notation, we say that the mixed
 strategies characterized by the probabilities $(p^{\star},q^{\star})$
 correspond to a Nash Equilibrium if the expected payoff functions satisfy
 the following conditions: 

\begin{equation}
\label{mixednash}
\left\{
 \begin{array}{ll}
 \bar{\$}_{1}(p^{\star},q^{\star}) \geq  
 \bar{\$}_{1}(p,q^{\star}) \:\:\:\:\:\: \forall\:\:p\in\:\: [0,1],
\\
\\
 \bar{\$}_{2}(p^{\star},q^{\star}) \geq  
 \bar{\$}_{2}(p^{\star},q) \:\:\:\:\:\: \forall\:\:q\in\:\: [0,1].
\end{array}
\right.
\end{equation} \\

We are now equipped with a more powerful mathematical tool, which can be
 applied to the analysis of the Battle of the Sexes.
 Let us suppose that Alice decides to resort to a probabilistic strategy in
which she chooses Opera ($\,s_{1}^{1}\,$) with probability $p$ ($\,$and
 obviously Television ($\,s_{1}^{2}\,$) with probabilty $1-p$$\,$) and Bob
 does the same, but with probability $q$ for Opera ($\,s_{2}^{1}\,$) and
 $1-q$ for Television ($\,s_{2}^{2}\,$).

We can then calculate their expected payoff functions through equation
(\ref{expected}), i.e. by summing all the
 joined probabilities relative to each couple of strategies, multiplied by
 the proper coefficients which appear in the bimatrix of the game, obtaining:

\begin{equation}
\bar{\$}_{A}(p,q)= p[q(\alpha -2\gamma +\beta) + \gamma -\beta] + \beta +
q(\gamma - \beta),
\end{equation}
\[ 
\bar{\$}_{B}(p,q)= q[p(\alpha -2\gamma +\beta) + \gamma -\alpha] + \alpha +
p(\gamma - \alpha).
\] 

Nash Equilibria can be found by imposing the two conditions:

\begin{equation}
\label{battlenash}
\left\{
 \begin{array}{ll}
 \bar{\$}_{A}(p^{\star},q^{\star})-\bar{\$}_{A}(p,q^{\star})=
 (p^{\star}-p)[\,q^{\star}(\alpha +\beta -2\gamma) -\beta +\gamma \,]
\geq 0 \:\:\:\:\:\:\forall\:\:p\in\:\: [0,1],
\\
\\
\bar{\$}_{B}(p^{\star},q^{\star})-\bar{\$}_{B}(p^{\star},q)=
 (q^{\star}-q)[\,p^{\star}(\alpha +\beta -2\gamma) -\alpha +\gamma \,]
\geq 0 \:\:\:\:\:\:\forall\:\:q\in\:\: [0,1].
\end{array}
\right.
\end{equation} 

They are satisfied if the two factors in each inequality have both
the same sign. There are three possibilities:

\begin{enumerate}
\item $p^{\star}_{(1)}=1,q^{\star}_{(1)}=1$.\\
In such a case both Alice and Bob choose to play $O$, i.e. a pure strategy.
Their corresponding expected payoff functions are $\bar{\$}_{A}(1,1)=\alpha$
 and $\bar{\$}_{B}(1,1)=\beta$, and since $\alpha > \beta > \gamma $ the
 following inequalities hold:

\begin{equation}
\left\{
 \begin{array}{ll}
\bar{\$}_{A}(1,1)-\bar{\$}_{A}(p,1)=(1-p)(\alpha - \gamma) \geq 0
\:\:\:\:\:\:\forall\:\:p\in\:\: [0,1],
\\
\\
\bar{\$}_{B}(1,1)-\bar{\$}_{B}(1,q)=(1-q)(\beta - \gamma) \geq 0
\:\:\:\:\:\:\forall\:\:q\in\:\: [0,1].
\end{array}
\right.
\end{equation}
 
\item $p^{\star}_{(2)}=0,q^{\star}_{(2)}=0$.\\
 The strategy is again a pure one, both Alice and Bob play $T$, and 
 $\bar{\$}_{A}(0,0)=\beta$ and
 $\bar{\$}_{B}(0,0)=\alpha$ ($\,$note that $\bar{\$}_{A}$ and $\bar{\$}_{B}$
 are reversed with respect to the previous case$\,$). Moreover:

\begin{equation}
\left\{
 \begin{array}{ll}
\bar{\$}_{A}(0,0)-\bar{\$}_{A}(p,0)=p(\beta - \gamma) \geq 0
\:\:\:\:\:\:\forall\:\:p\in\:\: [0,1],
\\
\\
\bar{\$}_{B}(0,0)-\bar{\$}_{B}(0,q)=q(\alpha - \gamma) \geq 0
\:\:\:\:\:\:\forall\:\:q\in\:\: [0,1].
\end{array}
\right.
\end{equation}
 
These two Nash Equilibria correspond to the solutions already found using pure
strategies only, but the introduction of mixed strategies allows for the
 existence of a third kind of solutions.

\item Inequalities (\ref{battlenash}) can be satisified also for $p^{\star}$
 and $q^{\star}$ different from $0$ or $1$. In such a case the factors
$p^{\star}-p$ and $q^{\star}-q$ can be positive or negative, depending on the
 values of $p$ and $q$. Then, the only way in order to fulfil 
conditions (\ref{battlenash}) consists in equating to zero the coefficients of 
$p^{\star}-p$ and $q^{\star}-q$, obtaining:

\begin{equation}
\label{terzanash}
p^{\star}_{(3)}=\frac{\alpha -\gamma}{\alpha +\beta -2\gamma},
\:\:\:\:\:\:\:\:\:\:
q^{\star}_{(3)}=\frac{\beta -\gamma}{\alpha +\beta -2\gamma}\:,
\end{equation}
 
where $p^{\star}$ and $q^{\star}$ are correctly larger than zero and less
than one.
Moreover they correspond to a Nash Equilibrium for which the gains of both
 players coincide:

\begin{equation}
\label{payoff}
\bar{\$}_{A}(p^{\star}_{(3)},q^{\star}_{(3)})=\bar{\$}_{B}(p^{\star}_{(3)},
q^{\star}_{(3)})= \frac{\alpha\beta  -{\gamma}^{2}}{\alpha +\beta -2\gamma}.
\end{equation}
 
 It can be easily shown that the expected payoff functions for both
players give now strictly less reward than the other two feasible strategies
$(O,O)$ or $(T,T)$, since: 

\begin{equation}
\gamma < \bar{\$}_{A(B)}(p^{\star}_{(3)},q^{\star}_{(3)})<\beta < \alpha.
\end{equation}

\end{enumerate}

We note that the solution of the game corresponding to the third case gives
 a good example of the exact meaning of Nash Equilibrium.
In fact, this peculiar concept of equilibrium can be defined only if all
 players are assumed to accept the same logic, i.e. that of achieving the
 best result if all the players play together at their best.
In the last case analysed the equilibrium corresponds to the conditions   
$\bar{\$}_{A}(p^{\star}_{(3)},q^{\star}_{(3)})=\bar{\$}_{A}(p,q^{\star}_{(3)})$
 and $\bar{\$}_{B}(p^{\star}_{(3)},q^{\star}_{(3)})=
\bar{\$}_{A}(p^{\star}_{(3)},q)$.
This means that if, for example, Alice accepts the logic of Nash Equilibrium
but Bob does not, the last one could choose the value of q minimizing
$\bar{\$}_{A}(p^{\star}_{(3)},q)$ without suffering any loss of gain.   

The introduction of mixed strategies has not solved the problem of Alice and 
Bob:
 even in this case the theory fails to say which one of the obtained Nash
 Equilibria represents the real development of the game, because they all
 satisfy our notion of rational solution and are therefore equally compelling.

We might now ask if it would be possible to obtain
a unique solution for the problem of the Battle of the Sexes by modifying again
 the theoretical structure of the game, in resemblance to what one does when 
 enriching the concept of pure strategy by defining a probability
 distribution over the strategic space.
This task is feasible introducing the notion of {\bf Quantum Strategy},
in connection with the concept of Entanglement.


\section{Quantum Game Theory}

The purpose of this section is that of extending the notion of a classical 
($\,$pure or mixed$\,$) strategy by giving a richer structure to the
strategic space, so that it contains the old set of strategies
as a proper subspace.
Therefore from now on, instead of considering only a discrete and finite set of
 strategies, we will permit the existence of linear superpositions of them,
 by giving the formal structure of an Hilbert space to the strategic space.
This allows the possibility of obtaining the {\it Pure Quantum Strategies}, 
defined as linear combinations, with complex coefficients, of pure
 classical ones.
According to the orthodox intepretation of quantum mechanics, the squared
modula of these complex coefficients have to be interpreted as the probability
of having played one particular pure classical strategy.
It is worthwhile observing that this intepretation of
 pure quantum strategies does not differ from the classical notion of a mixed
 strategy introduced before, because we are presently dealing with a
 restricted class of games ($\,$static games$\,$), where typical quantum
 interference effects between amplitudes do not occur.

Let us now apply the quantum approach
 to the study of the Battle of the Sexes, assuming that the
new space of strategies available to Alice and Bob consists of a
 two-dimensional Hilbert space, whose orthonormal basis vectors 
 $\vert O \rangle$ and $ \vert T \rangle$ correspond to classical strategies O 
and T, respectively.

An arbitrary pure quantum strategy is therefore described by the normalized
 state vector:

\begin{equation}
\vert \psi \rangle = a\, \vert  O \rangle + b\, \vert T \rangle\:\:,
\:\:\:\:\:\:\:\:\:\:\:\:\:\:\:\:\:\:aa^{\star} + bb^{\star} = 1.
\end{equation}

In order to use the normal form representation, we must
 define the quantum analogues of classical expected 
payoff functions and to decide the way each player can pick out his {\em best}
 strategy, i.e. the strategy which fulfils the conditions to be a Nash
 Equilibrium, if it does exist.
With the aim of solving these issues, we must make the following general
assumptions, which we hold to represent the common
 theoretical ingredients of all quantum games we can devise.

\begin{enumerate}

\item Unlike in the classical game theory, we have to fix an arbitrary initial
 quantum state belonging to the Hilbert space
 $\cal{S}={\cal{S}}_{A}\otimes {\cal{S}}_{B}$,
 obtained as a direct product of the two strategic spaces of the two players,
 whose orthonormal bases consist of the vectors associated to the pure
 classical strategies.

\item Each player can manipulate the initial state vector
 ($\,$his strategy$\,$) by performing some local transformation in order
to obtain a {\em suitable} final state vector which will represent the
quantum strategy the two players are going to play.
In this approach, since the states correspond to the strategy, the operators
 related to the transformations correspond to the tactics to be employed
in the game.

\item The expected payoff of each player must be evaluated by calculating
 first the squared modula of the projections of the final quantum state onto
 the basis vectors of space $\cal{S}$, and then by adding up the obtained
 numbers multiplied by the appropriate payoff coefficients ($\,$deducible from
 the payoff bimatrix$\,$).

\item Each player must eventually play the classical pure strategy, which
 results from a measurement process on the final quantum strategy, i.e.
 from a projection onto the
 canonical basis of his own strategic space.

\end{enumerate}

Before proceeding on, it is worthwhile making some comments about these 
assumptions, in order to make clear their meaning and utility.

Having given an Hilbert structure to the strategic space of each player, it is 
natural to describe their mutual choices as state vectors belonging to the
 direct product of their Hilbert spaces, which includes a continuous number
 of possible strategies.
Indeed, it is due to this richer structure that the Quantum Theory of Games
 exhibits more attractive features than its classical version: we will in
 fact be able to construct {\it factorizable strategies}, which are obtained
 as a direct product of two well-defined strategies for each single player,
 and {\it entangled strategies}, which cannot be decomposed in such a way and
 cannot be reduced to the previous ones by applying local transformations only.
 It will be seen that the presence of such an high degree of non-local 
 correlations between the strategies of the two players will provide
 us with a unique Nash Equilibrium solution of the Battle of The Sexes.
 
As stated in the first of the previous rules, we have introduced the notion
 of the initial quantum strategy for the two players: this state has no
 classical analogue and no particular meaning in our quantum context,
 representing only a starting vector which the two players must manipulate
in a suitable way in order to obtain the final strategy, this last one being
the real important object.
 
 If we start working with a factorizable initial state vector, the results
 ($\,$existence of Nash equilibria and value of expected payoff functions$\,$)
 of our treatment will not depend on its particular form, for each player can
 modify it with local operations in order to choose his own best strategy.
 Indeed, what is important in determining the reward given to each player,
 it is the form of the final state vector, which corresponds to the strategy
 effectively played by them\footnote{These considerations cannot be applied
 to reference~\cite{ref2}, where the results appear to be strictly initial
 state dependent.}.
This independence from the starting point  will not hold true, as we will see,
 in the case of entangled strategies, forcing us to impose limitations on the
 form of the initial vector. 

The second assumption deals with the possibility given to each player to
properly manipulate the initial state vector in order to find out the best
 final strategy, i.e. the one which satisfies the condition of corresponding
to a Nash Equilibrium.
The local transformations involved will depend on a certain number of
 parameters appearing as coefficients of the final strategy and
 determining the value of expected payoff functions.
 These parameters are the quantum analogues of the $p$ and $q$ parameters
 encountered in the classical mixed strategies, and the goal of each player
 will consist in searching for their best values.
 
The choice of the operators to be used is constrained only by the demand that,
 when using factorizable quantum states, the quantum version of the theory of
 games should be able to reproduce exactly all the results of the classical
 theory, whenever the players use mixed classical strategies.
We are therefore free to choose the operators, satisfying the forementioned
 constraint, in a such a way to make as simple as possible the calculations.

The third assumption deals with the probabilistic structure of the theory,
and asserts that each player can calculate an expected payoff function,
representing the average reward each player would receive if the game were
repeated many times ($\,$it has exactly the same meaning as in classical
theory$\,$), starting from the same final strategy.

Finally, the fourth assumption states that each player must effectively play
 the pure classical strategy which emerges after a process of measurement,
to be performed on the state vector representing the compound final strategy.
In fact, according to the usual postulates of Quantum Mechanics, after the
 measurement of a suitable observable, the final state vector collapses onto
 one of the vectors of the canonical basis of the strategic space ($\,$which
 are pure classical strategies, due to the first assumption$\,$).

 
\section{Factorizable Quantum Strategies}

Let us define the four-dimensional Hilbert space of common strategies
$\cal{S}={\cal{S}}_{A}\otimes{\cal{S}}_{B}$ of Alice and Bob for the Battle
 of the Sexes Game by giving its orthonormal basis vectors:

\begin{equation}
{\cal{S}}={\cal{S}}_{A}\otimes{\cal{S}}_{B}= ( \vert OO \rangle, \vert OT
 \rangle,\vert TO \rangle,\vert TT \rangle), 
\end{equation}

where the first position is reserved to the state of Alice and the second one
to that of Bob.
These vectors permit it to write all the possible pure classical strategies 
which Alice and Bob can simultaneously play, i.e. states of the type
$[\, x\vert 0 \rangle +y\vert T\rangle\,]\otimes [\,w\vert O \rangle +
z\vert T\rangle\,]$. 

Let us call $A$ and $B$ two unitary and unimodular matrices, representing
 the transformations that Alice and Bob may respectively use to manipulate
 their own strategies ($\,$they are written in the canonical basis where the
 first vector is denoted by $\vert O \rangle$ and the second by
 $\vert T \rangle$$\,$):

\begin{equation}
\label{operators}
 A= \left[ \begin{array}{cc}
        a         &  b \\
       -b^{\star} &  a^{\star} 
       \end{array}
\right], \:\:\:\:\:\:\:\:
 B= \left[ \begin{array}{cc}
        c         &  d \\
       -d^{\star} &  c^{\star} 
       \end{array}
\right],\:\:\:\:\:\:\:\:\:
\left\{ \begin{array}{l}
        aa^{\star}+bb^{\star}=1 \\
        cc^{\star}+dd^{\star}=1 
       \end{array}
\right. .
\end{equation}

Since a state of the type $r\vert O\rangle +s \vert T\rangle$ can always
be obtained from states $\vert O\rangle$ or $\vert T\rangle$ through $SU(2)$
transformations, the present procedure will be invariant with respect to the
 choice of the initial state.
 Then, we can make the calculation particularly simple starting from one of 
 the four states $\vert OO \rangle,\vert OT \rangle,\vert TO \rangle,\vert TT
 \rangle$.
Let us then choose for example $\vert \psi_{in}\rangle =\vert OO\rangle$.

By applying operators (\ref{operators}) to the initial state, we obtain a final
 common quantum strategy depending on the four complex coefficients $a,b,c,d$:

\begin{equation}
\label{psifinale}
\vert \psi_{fin} \rangle = A\otimes B\, \vert \psi_{in} \rangle 
= ac \vert OO \rangle -ad^{\star} \vert OT \rangle -
b^{\star}c \vert TO \rangle + b^{\star}d^{\star}  \vert TT \rangle. 
\end{equation}

 It is now possible to compute the expected payoff functions of both players
by projecting the final state onto the canonical basis of
 $\cal{S}={\cal{S}}_{A}\otimes{\cal{S}}_{B}$ and applying the third quantum
 rule:

\begin{equation}
\label{payofffinale}
\bar{\$}_{A}=\vert a \vert^{2} [(\alpha +\beta -2\gamma) \vert c \vert^{2}
 -\beta+\gamma ] +\beta +(\gamma - \beta) \vert c \vert^{2},
\end{equation}
\[
\bar{\$}_{B}= \vert c \vert^{2} [ (\alpha +\beta -2\gamma) \vert a \vert^{2} -
\alpha + \gamma] + \alpha +(\gamma -\alpha)\vert a \vert^{2}, 
\]
  
where we have eliminated the parameters $b$ and $d$.

Now Alice and Bob can respectively find out which are the values of
 coefficients $\vert a \vert^{2}$ and  $\vert c \vert^{2}$ satisfying
the Nash Equilibrium condition, whose defining inequalities (3.2) remain
 valid also for Quantum Game Theory.
By repeating calculations similar to those of section 3, it is easy to
 notice that the desired values for the coefficients are
 represented by the following pairs:

\begin{equation}
(\vert a \vert^{2}=0\, , \, \vert c \vert^{2}=0) \:\:\:or\:\:\:
(\vert a \vert^{2}=1\, , \, \vert c \vert^{2}=1) \:\:\:or\:\:\: 
(\vert a \vert^{2}=\frac{\alpha -\gamma}{\alpha +\beta -2\gamma}\, , \,
 \vert c \vert^{2}=\frac{\beta -\gamma}{\alpha +\beta -2\gamma})
\end{equation}

The first and the second pair of values are the quantum analogues of the pure
classical strategies in which both players choose the same strategy. The final
quantum strategy and the relative expected payoff functions are easily 
obtained by substituting these values in equations (\ref{psifinale}) and
(\ref{payofffinale}):

\begin{equation}
\label{cinquesei}
(\vert a \vert^{2}=0\, , \, \vert c \vert^{2}=0)
\:\:\:\:\: \Rightarrow\:\:\:\:\:
\vert \psi_{fin} \rangle = \vert TT \rangle
 \:\:\:\:\: \Rightarrow\:\:\:\:\: 
\bar{\$}_{A}=\beta \:\:\:\:\bar{\$}_{B}=\alpha, 
\end{equation}
\[
(\vert a \vert^{2}=1\, , \, \vert c \vert^{2}=1)
\:\:\:\:\: \Rightarrow\:\:\:\:\:
\vert \psi_{fin} \rangle = \vert OO \rangle
 \:\:\:\:\: \Rightarrow\:\:\:\:\: 
\bar{\$}_{A}=\alpha \:\:\:\:\bar{\$}_{B}=\beta.
\] 

The third pair of values for coefficients $\vert a \vert^{2}$ and  
$\vert c \vert^{2}$ corresponds to a mixed classical strategy played with
 probabilities given by equation (\ref{terzanash}).
In fact, the factorizable final quantum strategy\footnote{We have omitted for
simplicity useless phase factors in the coefficients of the final states, since
 we are considering only static games, where there is no evolution.} and the
expected payoff functions for Alice and Bob assume the following form:

\begin{equation}
\label{psifinalemixed}
\vert \psi_{fin} \rangle = \frac{1}{\alpha +\beta -2\gamma}\,
[\,\sqrt{\alpha -\gamma} \vert O \rangle -\sqrt{\beta -
\gamma} \vert T \rangle\,]\cdot
[\, \sqrt{\beta -\gamma} \vert O \rangle -\sqrt{\alpha -
\gamma} \vert T \rangle\,] \\
\end{equation}
\[
\Rightarrow\:\:\:\:\: 
\bar{\$}_{A}=\bar{\$}_{B}=\frac{\alpha\beta -{\gamma}^{2}}{\alpha +\beta
 -2\gamma}\:.
\]

The quantities $\bar{\$}_{A}$ and $\bar{\$}_{B}$ have the same values we found
in the case of mixed classical strategies -- see equation (\ref{payoff}).

Due to the forementioned invariance, if we were started from another initial
 quantum state we would have obtained different values for coefficients
$\vert a \vert^{2}$ and $\vert c \vert^{2}$ satisfying Nash Equilibrium
 conditions, but the corresponding expressions for the final quantum strategy
 and the expected payoff functions for both players would be unchanged
 ($\,$always apart from phase factors$\,$).

Concluding, we see that factorizable quantum strategies are able to reproduce
 exactly the same results obtained before, when dealing with the mixed
 classical version of the Battle of the Sexes game.
 The exploitation of a quantum formalism and, particularly, of factorizable
 strategies does not constitute an improvement with respect to the
 classical theory: the game remains undecidible, since there is no way to
 prefer one Nash strategy with respect to the other ones. Nevertheless, 
 Quantum Game Theory contains the Classical one as a subset, and quantum
 factorizable strategies correspond to classical mixed ones.  


\section{Density Matrix Approach to Quantum Game Theory}

Before beginning the study of the properties of entangled quantum strategies,
we show how to rewrite the entire formalism developed so far
using density matrices instead of state vectors, and introducing a new kind of
 transformations.
It is worthwhile doing it in order to be able to make simple the mathematical
 calculations when entangled strategies are used, bearing in mind that, in the
 case of factorizable states, the solutions we got in the previous section
 must be obtained again.

Let us therefore define a unitary and hermitian operator $C$ interchanging 
vectors $\vert O \rangle$ and $\vert T \rangle $:

\begin{equation}
\left\{
\begin{array}{l}
  C\,\vert O \rangle = \vert T \rangle \\
  C\,\vert T \rangle = \vert O \rangle 
\end{array}
\right.
\:\:\:\:,\:\:\:\:\:C^{\dagger}=C=C^{-1} \:.
\end{equation} 

We assume now that each player can modify his own strategy by applying
to his reduced part of the total density matrix $\rho_{in}$, which represents
 the initial state of the game, the following transformation:

 \begin{equation}    
\label{convex}
\rho_{fin}^{A(B)}=[\, p\,I\,\rho^{A(B)}_{in}\,I^{\dagger} + (1-p)\,C \,
\rho^{A(B)}_{in}\, C^{\dagger}\,]\:.  
\end{equation} 
 
This operation can be interpreted as the choice of each player to act
 with the identity $I$ with probability $p$ and with $C$ with probability 
$(1-p)$, 
 and gives rise to the following final density matrix:

\begin{eqnarray}
\label{rhofinale}
\rho_{fin} & = & pq\,
 I_{A}\otimes I_{B}\, \rho_{in}\, I^{\dagger}_{A}\otimes I_{B}^{\dagger} +
 p(1-q)I_{A}\otimes C_{B}\, \rho_{in}\, I^{\dagger}_{A}\otimes C_{B}^{\dagger}+
 \nonumber \\
 & &  q(1-p) C_{A}\otimes I_{B}\, \rho_{in}\, C_{A}^{\dagger}\otimes
 I_{B}^{\dagger}+ (1-p)(1-q) C_{A}\otimes C_{B}\,\rho_{in}\, C_{A}^{\dagger}
\otimes C_{B}^{\dagger}
\end{eqnarray}

It is possible to show that, if we start from a density matrix corresponding to
 one of the four states $\vert OO \rangle,\vert OT \rangle,\vert TO \rangle$
 and $\vert TT \rangle$ which we could have been used indifferently in the case
 of factorizable quantum strategies, the particular kind of transformation
 employed in equation (\ref{convex}) reproduces the same results obtained in
 section $5$.

Let us choose for example the initial density matrix corresponding to the state
$\vert \psi_{in}\rangle=\vert OO\rangle$ used in the preceding section.
Then from equation (\ref{rhofinale}) we obtain:

\begin{eqnarray}
\rho_{fin} & = & pq \vert OO\rangle \langle OO \vert + p(1-q)\vert OT\rangle 
\langle
OT \vert + \nonumber \\
& & (1-p)q\vert TO\rangle \langle TO \vert +(1-p)(1-q)\vert TT\rangle \langle T
T \vert.
\end{eqnarray}
  
In order to calculate the payoff functions, we must introduce two payoff
 operators:

\begin{equation}
P_{A}=\alpha \vert OO \rangle\langle OO \vert +
 \gamma(\vert OT \rangle\langle OT \vert +\vert TO \rangle\langle TO \vert)
+\beta \vert TT \rangle\langle TT\vert, 
\end{equation}
\[ P_{B}=\beta \vert OO \rangle\langle OO \vert +
 \gamma(\vert OT \rangle\langle OT \vert +\vert TO \rangle\langle TO \vert)
+\alpha \vert TT \rangle\langle TT\vert.
\]

The payoff functions can then be obtained as mean values of these operators:

\begin{equation}
\bar{\$}_{A}=Tr(P_{A}\,\rho_{fin})\:\:\:\:\:\:
\bar{\$}_{B}=Tr(P_{B}\,\rho_{fin}).
\end{equation}

 Three Nash Equilibria arise: those
 for $(p^{\star}=1,q^{\star}=1)$ and for $(p^{\star}=0,q^{\star}=0)$, 
corresponding to
 $\rho_{fin}=\vert OO\rangle\langle OO \vert $ and $\rho_{fin}=\vert 
TT\rangle\langle
 TT \vert$, respectively, and a third one associated to the final density 
matrix:

\begin{eqnarray}
\label{seisette}
\rho_{fin} & = & \frac{1}{\alpha+\beta -2\gamma}[\, (\alpha -\gamma)
\vert O\rangle\langle O\vert + (\beta -\gamma)\vert T\rangle\langle T\vert\,]
\otimes \nonumber\\
& & \frac{1}{\alpha+\beta -2\gamma}[\, (\beta -\gamma)
\vert O\rangle\langle O\vert + (\alpha -\gamma)\vert T\rangle\langle T\vert\,]
\end{eqnarray}

The payoff functions corresponding to these strategies are the same we have
already obtained in the previous section -- see equations (\ref{cinquesei})
 and (\ref{psifinalemixed})\footnote{We note that starting with the state
 $\vert TT\rangle$ one obtains the same Nash Equilibria but now
 $(p^{\star}=1,q^{\star}=1)$ corresponds to
 $\rho_{fin}=\vert TT\rangle\langle TT\vert$ and $(p^{\star}=0,q^{\star}=0)$
 to $\rho_{fin}=\vert OO\rangle\langle OO\vert$.
If we choose $\vert OT \rangle$ or $\vert TO \rangle$ as starting states,
 the same final density matrices are found for $(p^{\star}=0,q^{\star}=1)$ and
$(p^{\star}=1,q^{\star}=0)$ or $(p^{\star}=1,q^{\star}=0)$ and 
$(p^{\star}=0,q^{\star}=1)$, respectively.
The third Nash Equilibrium corresponding to the $\rho_{fin}$ given by
 (\ref{seisette}) is always found.}

Concluding, we have shown that it is possible to get the same results one 
obtains in the classical version of our game, by allowing the two players
 to manipulate their own strategy with unitary and unimodular operators
 or, equivalently, through a particular transformation involving two hermitian
 operators ($\,$$C$ and $I$$\,$), one interchanging states $\vert O\rangle$
 and $\vert T\rangle$ and the other leaving them unvaried.


\section{Entangled Quantum Strategies}

In sections 5 and 6 we have seen, using two different but equivalent ways, 
that the exploitation of factorizable quantum strategies does not modify at
 all the results we can obtain by applying classical game theory to the
 analysis of the Battle of the Sexes game.
 On the other hand, it is just the richer structure we have given to the
 strategic space of both players that allows us to obtain new results.
In fact, up to here we have not introduced entangled states in the quantum
 version of the game: as a matter of fact, we are going to show how an
 entangled couple of strategies played by Alice and Bob will lead to a
 unique solution of the game, always satisfying the Nash Equilibrium
 condition. 

According to what as been shown in the previous section, we can restrict
ourselves to deal with density matrices associated to strategies, rather
than to state vectors, and to use the transformations of equation
 (\ref{rhofinale}).
The choice of the initial density matrix is relevant, in so far as the payoff 
functions corresponding to the Nash Equilibria turn out to be all initial
 state dependent. 

Let us start assuming that Alice and Bob have at their disposal the following
 entangled state:

\begin{equation}
\vert \psi_{in} \rangle = a\vert OO\rangle +b\vert TT\rangle\:\:\:\:\:\:\:
\vert a \vert^{2}+\vert b \vert^{2}=1\, ,
\end{equation}

with the associated density matrix:

\begin{equation} 
\rho_{in}= \vert a \vert^{2} \vert OO \rangle \langle OO \vert +
ab^{\star}\vert OO \rangle \langle TT \vert +
a^{\star}b\vert TT \rangle \langle OO \vert +
\vert b \vert^{2}\vert TT \rangle \langle TT \vert.
\end{equation}

By using equation (\ref{rhofinale}) we obtain
the expression for the final density matrix $\rho_{fin}$, which depends on
parameters $a,b,p$ and $q$, and, through eq. (6.7), also the expected payoff 
functions for both players:

\begin{eqnarray}
\bar{\$}_{A}(p,q) & = & p[\, q(\alpha+\beta-2\gamma) -\alpha\vert b \vert^{2}
-\beta\vert a \vert^{2}+\gamma\,] + \nonumber \\
 & & q[\,-\alpha\vert b \vert^{2} -\beta\vert a \vert^{2}+\gamma\,] +
 \alpha\vert b \vert^{2}+\beta\vert a \vert^{2},\nonumber \\
\bar{\$}_{B}(p,q) & = & q[\, p(\alpha+\beta-2\gamma) -\beta\vert b \vert^{2}
-\alpha\vert a \vert^{2}+\gamma\,] +\nonumber \\
 & & p[\,-\beta\vert b \vert^{2} -\alpha\vert a \vert^{2}+\gamma\,] +
 \beta\vert b \vert^{2}+\alpha\vert a \vert^{2}.
\end{eqnarray}

Then, three Nash Equilibria arise from the following two
inequalities:

\begin{equation}
\left\{
  \begin{array}{ll}
\bar{\$}_{A}(p^{\star},q^{\star})-\bar{\$}_{A}(p,q^{\star})=
(p^{\star}-p)[\, q^{\star}(\alpha+\beta-2\gamma) -\alpha\vert b \vert^{2}
-\beta\vert a \vert^{2}+\gamma\,] \geq 0,
\\
\bar{\$}_{B}(p^{\star},q^{\star})-\bar{\$}_{B}(p^{\star},q)=
(q^{\star}-q)[\, p^{\star}(\alpha+\beta-2\gamma) -\beta\vert b \vert^{2}
-\alpha\vert a \vert^{2}+\gamma\,] \geq 0.
\end{array}
\right.
\end{equation}

Let us examine separately the three different Nash Equilibria:

\begin{enumerate}
\item $p^{\star}_{(1)}=q^{\star}_{(1)}=1$.\\
In such a case, the initial and the final density matrix representing
the common strategies played by the players turn out to be equal and the 
expected 
payoff functions are:

\begin{eqnarray}
\bar{\$}_{A}(1,1)& =&\alpha\vert a \vert^{2}+
 \beta\vert b \vert^{2}\\
\bar{\$}_{B}(1,1)& =&\beta\vert a \vert^{2}+
 \alpha\vert b \vert^{2}
\end{eqnarray}

\item $p^{\star}_{(2)}=q^{\star}_{(2)}=0$.\\
In correspondence to these values the final density
matrix is obtained by reversing the initial strategies ($\,$$\rho_{fin}=
C_{A}C_{B}\rho_{in}C^{\dagger}_{B}C^{\dagger}_{A}$$\,$), and the corresponding
 expected payoff functions turn out to be interchanged with respect to
 the previous case:

\begin{eqnarray}
\bar{\$}_{A}(0,0)& =&\beta\vert a \vert^{2}+
 \alpha\vert b \vert^{2},\\
\bar{\$}_{B}(0,0)& =&\alpha\vert a \vert^{2}+
 \beta\vert b \vert^{2}
\end{eqnarray}
 
\item $p^{\star}_{(3)}=\frac{(\alpha-\gamma)\vert a \vert^{2}+
(\beta-\gamma)\vert b \vert^{2}}{\alpha+\beta-2\gamma},\:\:
q^{\star}_{(3)}=\frac{(\alpha-\gamma)\vert b \vert^{2}+
(\beta-\gamma)\vert a \vert^{2}}{\alpha+\beta-2\gamma}.$\\

Due to the condition $\alpha>\beta>\gamma$ one sees immediately that 
$0<p^{\star}_{(3)}<1$ and $0<q^{\star}_{(3)}<1$.
The expected payoff functions of both players turn out to be the same,
and display the following dependance on the parameters of the game:

\begin{equation}
\bar{\$}_{A}(p^{\star}_{(3)},q^{\star}_{(3)})=
\bar{\$}_{B}(p^{\star}_{(3)},q^{\star}_{(3)})=
\frac{1}{\alpha +\beta -2\gamma}[\, \alpha\beta  +( \alpha -\beta)^{2}\vert
a \vert^{2}\vert b \vert^{2}
-\gamma^{2}\,]
\end{equation}

\end{enumerate}

We have found again three different Nash Equilibria for the game, whose
expected payoff functions are so related:

\begin{equation}
\bar{\$}_{A(B)}(p^{\star}_{(3)},q^{\star}_{(3)})<\bar{\$}_{A(B)}(p^{\star}
_{(1)},q^{\star}_{(1)})\:\:\:\:,\:\:\:\:
\bar{\$}_{A(B)}(p^{\star}_{(3)},q^{\star}_{(3)})<\bar{\$}_{A(B)}(p^{\star}
_{(2)},q^{\star}_{(2)}).
\end{equation}

This simply means that both Alice and Bob would prefer to play strategies
$p^{\star}_{(1)}=q^{\star}_{(1)}=1$ or $p^{\star}_{(2)}=q^{\star}_{(2)}=0$
 rather than the third one, nonetheless being unable to decide which one of
 the two. In fact:

\begin{eqnarray}
\bar{\$}_{A}(1,1)-\bar{\$}_{A}(0,0)& =&(\alpha-\beta)(\vert a \vert^{2}-
\vert b \vert^{2}), \nonumber\\
\bar{\$}_{B}(1,1)-\bar{\$}_{B}(0,0)& =&(\alpha-\beta)(\vert b \vert^{2}-
\vert a \vert^{2}).
\end{eqnarray}

Therefore, for $\vert a \vert>\vert b \vert$ Alice would prefer the first Nash
 Equilibrium, while Bob prefers the second one, and the contrary happens if
 $\vert a \vert<\vert b \vert$.
However, there is a case in which both Alice and Bob can make the same
choice: it corresponds to starting with an initial state having 
$\vert a \vert=\vert b \vert$, i.e. with:

\begin{equation}
\vert \psi_{in} \rangle = \frac{1}{\sqrt{2}} \left[ \, \vert OO \rangle +
\vert TT \rangle \, \right],
\end{equation}
\[ \rho_{in}=\frac{1}{2} \left[ \, \vert OO \rangle \langle OO \vert +
\vert OO \rangle \langle TT \vert +\vert TT \rangle \langle OO \vert +
\vert TT \rangle \langle TT \vert \, \right]. 
\]

It is now trivial to find
out which are the Nash Equilibria for this peculiar initial state. We have:

\begin{eqnarray}
 (p^{\star}=0,q^{\star}=0) &  \Rightarrow &
 {\$}_{A}={\$}_{B}=\frac{(\alpha+\beta)}{2} \nonumber  \\
(p^{\star}=1,q^{\star}=1) & \Rightarrow &
{\$}_{A}={\$}_{B}=\frac{(\alpha+\beta)}{2} \nonumber  \\
 (p^{\star}=\frac{1}{2},q^{\star}=\frac{1}{2}) & \Rightarrow &
{\$}_{A}={\$}_{B}=\frac{\alpha+\beta+2\gamma}{4}
\end{eqnarray}

The fact that both players have the same expected payoff functions constitutes
 the most attractive feature of having exploited entangled strategies.
 In fact, now it is possible to choose a unique Nash Equilibrium by 
 discarding the ones which give the players the lesser reward: the remaining
 couple of values for $p$ and $q$ will then represent the unique solution of
 the game, which may satisfy both players, in contrast to what happens in the
 case of factorizable strategies, where there is no way to choose one of the
 three equivalent strategies. In fact, from eq.s (7.13) one sees immediately
 that:

\begin{equation}
\bar{\$}_{A(B)}(p^{\star}=0,q^{\star}=0)\, =\,\bar{\$}_{A(B)}(p^{\star}=1,
q^{\star}=1)\,
 > \bar{\$}_{A(B)}(p^{\star}=1/2,q^{\star}=1/2).
\end{equation}

The entangled quantum strategy which corresponds to both values 
$(p^{\star}=0,q^{\star}=0)$ and $(p^{\star}=1,q^{\star}=1)$ is the same and it
 is precisely the entangled state we started from:

\begin{equation}
\label{entangled}
\vert \psi_{fin} \rangle = \frac{1}{\sqrt{2}} \left[ \, \vert OO \rangle +
\vert TT \rangle \, \right].
\end{equation}

We have therefore obtained the result that the
 state (7.15) represents the quantum entangled pair of strategies
 which satisfy the Nash Equilibrium conditions, i.e. it represents the best
 rational couple of choices which are stable against unilateral
deviations, and gives an higher reward to both players than the other possible
 Nash Equilibrium $(p^{\star}=1/2,q^{\star}=1/2)$. 
 
The entangled strategy (7.15) can therefore be termed the unique
 solution of the quantum version of the Battle of the Sexes game.

\section{Conclusions}
 
In this paper the classical theory of games has been extended to a quantum
 domain, by giving an Hilbert structure to the strategic spaces of the
players, so allowing the existence of linear combinations of classical
 strategies to be interpreted according to the usual formalism of orthodox
quantum mechanics.
The only remarkable assumption we have made, reproducing the usual results
of the classical game theory in the case of factorizable quantum strategies,
is that each player must choose his strategy by performing quantum
 measurements on his state vector.
 Another weak assumption deals with the form of the operator which both
 players can apply in order to properly manipulate their strategies: we have
 shown that is sufficient, to reproduce the classical results and to
 obtain new ones, to restrict our attention to a transformation involving two
 particular hermitian operators.

We have applied this formalism to the study of the Battle of the Sexes game,
 in order to find out in which cases a quantum strategy could exhibit more
 attractive features than the classical theory predicts; it has been shown
 that the use of factorizable quantum strategies ($\,$i.e. factorizable
 states in the strategic space$\,$) by both players, cannot improve their
 expected payoff functions and reproduces the same behaviour of
 the classical version of the game: the Battle admits three Nash Equilibria
which can be considered as three possible solutions of the game and the
players cannot rationally decide which one to choose.
On the contrary, it has been shown  that if both players are allowed to play
 entangled quantum strategies, the game admits again three possible Nash
 Equilibria, but a unique solution. In fact, in this situation both players
 get the same expected payoff and the solution is represented by the Nash
 strategy which gives more reward. This new feature is originated from the
 fact that both players are forced to act in the same way, this
 being due to the strong correlation present in the entangled
form of the solution strategy $\frac{1}{\sqrt{2}} (\vert OO \rangle +
\vert TT \rangle)$. The meaning of this result is in some sense obvious. 
If Alice and Bob are obliged to play the same strategy and have the possibility
 of choosing together sometimes O and sometimes T ($\,$this peculiarity
 not being present in the classical game and in the factorizable quantum
 case$\,$), it is a priori clear that they obtain the best reward playing
 one-half times $O$ and one-half $T$.
In any case, what is important is the fact that the entangled strategies,
 in the context of a quantum approach to the Theory of Games, give such a
 result.

Concluding, we can observe that it could be interesting to apply the quantum
 version to the study of the dynamic games. In such a case the transformations
 applied to the strategies represent the moves played sequentially by the
 players. New results could then come out from some clever choice of these
 transformations.


\end{document}